\documentclass[twocolumn,showpacs,aps,floatfix,superscriptaddress]{revtex4}
\usepackage{amsmath,amssymb,graphicx,eucal}
\begin{document}
\title{Random Ancestor Trees}
\author{E.~Ben-Naim}
\affiliation{Theoretical Division and Center for Nonlinear
Studies, Los Alamos National Laboratory, Los Alamos, New Mexico
87545, USA}
\author{P.~L.~Krapivsky}
\affiliation{Department of Physics,
Boston University, Boston, Massachusetts 02215, USA}
\begin{abstract}
We investigate a network growth model in which the genealogy controls
the evolution. In this model, a new node selects a random target node
and links either to this target node, or to its parent, or to its
grandparent, etc; all nodes from the target node to its most ancient
ancestor are equiprobable destinations.  The emerging random ancestor
tree is very shallow: the fraction $g_n$ of nodes at distance $n$ from
the root decreases super-exponentially with $n$, $g_n=e^{-1}/(n-1)!$.
We find that a macroscopic hub at the root coexists with highly
connected nodes at higher generations. The maximal degree of a node at
the $n$th generation grows algebraically as $N^{1/\beta_n}$ where $N$
is the system size.  We obtain the series of nontrivial exponents
which are roots of transcendental equations: $\beta_1\cong 1.351746$,
$\beta_2\cong 1.682201$, etc.  As a consequence, the fraction $p_k$ of
nodes with degree $k$ has algebraic tail, $p_k\sim k^{-\gamma}$,
with $\gamma=\beta_1+1=2.351746$.
\end{abstract}
\pacs{89.75.Hc, 02.10.Ox, 05.40.-a, 02.50.Ey}
\maketitle

\section{Introduction}

The interplay between dynamics, structure, and function of complex
networks is the subject of intense current research
\cite{ws,ab,dm,bbv}. A wide spectrum of social, technical, and
biological networks have broad degree distributions with a power-law
tail \cite{ab,dm}. Further, many real-life networks also have
macroscopic hubs that are connected to a finite fraction of all nodes
and these hubs play important function in the network
\cite{ab,dm,bbv}.  In this study, we introduce a minimal model that
captures both of these features.

Common mechanisms of network growth include preferential attachment
\cite{yule,simon,ba,krl,dms}, copying, and re-direction
\cite{kum,kr,la,kr2}. In these processes, a target node is identified
and the attachment is either to the target node itself or to one of
its immediate ancestors.  These growth processes result in
heterogeneous networks with broad degree distributions.  However,
there are no macroscopic hubs because the maximal degree in these
networks grows sub-linearly with the network size.

In this study, we consider the complementary situation where
attachment to any ancestor is possible.  We show that the resulting
network has a fundamentally different structure: there is coexistence
between a hub with a macroscopic degree and highly connected nodes
with sub-macroscopic degrees.  Moreover, there is a series of
exponents that characterize how the maximal degree at a given distance
from the root scales with system size.

\begin{figure}[t]
\includegraphics[width=0.38\textwidth]{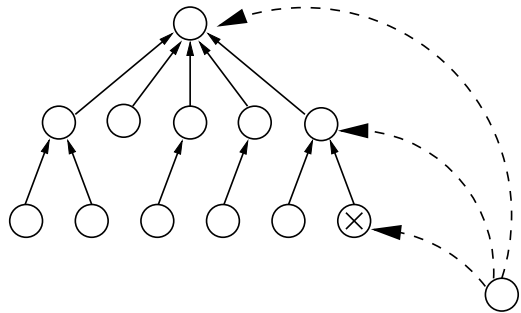}
\caption{The random ancestor tree. A new node attaches, with equal
probabilities, to one of three nodes: the target node (marked by
$\times$), its parent, or its grandparent.}
\label{fig-tree}
\end{figure}

To model situations where the genealogy controls the network
growth, we start with a single node, the root, and add nodes
sequentially, so that each new node attaches to an existing node
through a two-step process (see Fig.~\ref{fig-tree}):
\begin{itemize}
\item The new node selects a target node at random.
\item The new node attaches to one node, selected at random from the
set of nodes that includes the target node and all its ancestors.
\end{itemize}
Since each new node adds a single link, the growing network retains a
tree structure, and apart from the root, each node has a single parent
node.  We term the resulting structure the {\em random ancestor tree}.
An appealing feature of this model is the absence of control
parameters.

Our motivation for this model comes from social, scholarly, and
reference networks \cite{wf}. For example, the published literature in
science, law, and religion grows by sequential addition
of material. Starting with some reference, a scholar typically studies
an entire line of preceding reports or rulings, and refers to
the most appropriate or the most important published literature, not
necessarily the most recent one \cite{sr, mejn, rfmv}. In this sense, the
evolution of such reference networks is coupled to their genealogy.

We find that the random ancestor tree has a single macroscopic hub,
namely the root, that is connected to a fraction $e^{-1}$ of all
nodes. Furthermore, the degree distribution is broad as the fraction
$p_k$ of nodes with degree $k$ decays algebraically,
\begin{equation}
\label{announce}
p_k\sim k^{-\gamma},\qquad \gamma=2.351746,
\end{equation}
when $k\to\infty$. Interestingly, the exponent $\gamma$ is obtained as
a root of a transcendental equation. We also obtain a series of
nontrivial exponents $\gamma_n$, each characterizing the degree
distribution of nodes at the $n$th generation. As $n$ increases, the
exponents $\gamma_n$ increase, so that the nodes become less and less
connected as their distance from the root increases.

Throughout this paper, we focus on the asymptotic behavior of large
trees.  We first analyze the genealogical structure of the random
ancestor tree in section \ref{GT}.  The depth distribution indicates
that the root is a macroscopic hub, and in section \ref{hubs} we show
that the root is the only such hub. We then establish
(Sec.~\ref{high}) the degree of the most connected nodes at higher
generations and discuss the implications of these results for the
degree distribution (Sec.~\ref{degree}).  Other structural properties
including the number of leaves, and the likelihood that the tree has
star or chain topology are derived in section
\ref{leaves}. Conclusions are given in section \ref{conclude}.

\section{Genealogical Structure}
\label{GT}

We treat the random ancestor tree as a genealogical tree and group
nodes according to generation (figure \ref{fig-tree}). The first
generation consists of nodes at distance one from the root, the second
generation includes nodes at distance two from the root, etc.  In
principle, the number of nodes in each generation fluctuates from one
realization to another, yet in the thermodynamic limit such
fluctuations are negligible, and we therefore study the average number
of nodes in each generation.

Let $N$ be the total number of nodes in the tree and let $G_n(N)$ be
the average number of nodes at distance $n$ from the root. Only the
root belongs to the zeroth generation and therefore, $G_0(N)=1$. The
average number of nodes in the first generation, $G_1(N)$, grows according
to the rate equation
\begin{equation}
\label{G1}
\frac{dG_1}{dN} = \frac{G_0}{N} + \frac{G_1}{2N} + \frac{G_2}{3N} +\cdots.
\end{equation}
Hereinafter we treat the size of the network $N$ as a continuous
variable; our results are asymptotically exact in the limit
$N\to\infty$. Equation \eqref{G1} reflects that new nodes are added to
the first generation only when a new node links to the root. The first
term on the right-hand side corresponds to the situation when the root
is chosen as the target node; the second term accounts for situations
when a node in the first generation is chosen --- then the probability
that the actual link will be to the root is 1/2; the following terms
describe situations where the target nodes are in higher generations.

The generalization of \eqref{G1} to arbitrary generation $n$ is
straightforward,
\begin{equation}
\label{Gn}
\frac{dG_n}{dN} = \sum_{i\geq n} \frac{G_{i-1}}{i\,N}.
\end{equation}
We now introduce $g_n$, the fraction of nodes that belong to the $n$th
generation,
\begin{equation}
\label{Gn-seek}
G_n(N) = g_n N.
\end{equation}
The coefficients $g_n$ are normalized, $\sum_{n\geq 1}
g_n=1$. Further, since $G_0=1$ we have $g_0=0$ in the limit $N\to
\infty$.  Equation \eqref{Gn} shows that the fractions $g_n$ satisfy
\begin{equation}
\label{gn}
g_n = \sum_{i\geq n} i^{-1} g_{i-1}\,.
\end{equation}
It is convenient to re-write Eq.~\eqref{gn} as the explicit recursion
\begin{equation}
\label{gn-rec}
g_n= n^{-1} g_{n-1} + g_{n+1}\,.
\end{equation}
By evaluating the first few terms, $g_2=g_1$, $g_3=g_1/2!$,
$g_4=g_1/3!$, and by induction, $g_n=g_1/(n-1)!$.  The normalization
condition $\sum_{n\geq 1} g_n=1$ sets $g_1=e^{-1}$. Thus, 
\begin{equation}
\label{gn-sol}
g_n = \frac{e^{-1}}{(n-1)!}, 
\end{equation}
indicating that the number of nodes decays super-exponentially with
increasing generation or alternatively, depth. Hence, the random
ancestor tree is exceptionally shallow. For instance, the average
depth is finite: $\langle n\rangle=\sum_{n\geq 1} n g_n=2$.

The genealogical structure of the random ancestor tree drastically
differs from the genealogical structure of other growing networks
\cite{kr}. In particular, random recursive trees \cite{hmm,krb}, which
grow by attachment of new nodes to random target nodes, have the depth
distribution $G_n(N)=(\ln N)^n/n!$. Thus, there is a sharp peak at
$n_{\rm peak}\sim \ln N$, and furthermore, the maximal depth, defined
as the deepest non-empty generation $n_{\rm max}$, also scales
logarithmically with system size, $n_{\rm max}\sim \ln N$
\cite{kr}. This behavior is very robust and occurs in most growing
trees \cite{bp,nsw,bo}, but there are exceptions where the maximal
depth grows algebraically with system size \cite{dkm}.

For the random ancestor tree, the generation profile \eqref{gn-sol} is
a sharply decreasing function of depth, and from the criterion
$G_{n_{\rm max}}\sim 1$ we estimate the maximal depth
\begin{equation}
\label{n-max}
n_{\rm max} \approx \frac{\ln N}{\ln(\ln N)}\,.
\end{equation}
This depth behavior reflects that the tree is shallow \cite{bk}.

\section{Macroscopic Hub}
\label{hubs}

The random ancestor tree has a macroscopic hub: the root is connected
to a finite fraction, $e^{-1}=0.367879$, of all nodes as follows from
\eqref{gn-sol}. Such macroscopic hubs are found in social networks
\cite{wf}, and have an important function in as transportation
networks \cite{bbv,ab1,yb,gm,wrd}.  Yet, with a few exceptions
including super-linear preferential attachment \cite{krl,kr,kk}, most
models of network growth have no such hubs and instead, the most
central nodes have degrees that grow sub-linearly with network size.

To find out whether, in addition to the root, there are other hubs,
let's consider the most connected node at the first generation.  We
denote the number of descendants of this node in the second generation
by $H_2(N)$, the number of its descendants in the third generation by
$H_3(N)$, etc.  The equation governing $H_j(N)$ is identical to
\eqref{Gn},
\begin{equation}
\label{Hj}
\frac{dH_j}{dN} = \sum_{i\geq j} \frac{H_{i-1}}{i\,N}\,,
\end{equation}
with $H_1=1$. By definition, the node degree is $1+H_2(N)$.

Suppose that the most connected node at the first generation has
macroscopic degree, $H_2\sim N$. Since $dH_3/dN>H_2/(3N)$ then
$H_3\sim N$ as well. In general,
\begin{equation}
\label{Hj-seek1}
H_j(N) = u_j N
\end{equation}
for all $j\geq 2$.  By substituting \eqref{Hj-seek1} into \eqref{Hj},
we find that the coefficients $u_j$ satisfy a recursion relation
identical to \eqref{gn-rec},
\begin{equation}
\label{uj-rec}
u_j = j^{-1} u_{j-1} +u_{j+1}.
\end{equation}
Any solution to \eqref{uj-rec} is a linear combination of the two
independent solutions: the fast-decaying solution above $u_j=1/(j-1)!$
and the slow-decaying one, $u_j\sim j^{-1}$ when $j\gg 1$.  For the
first few coefficients we have $u_1=0$ and $u_2=u_3$. The boundary
condition $u_1=0$ implies that we can not express the solution solely
in terms of the rapidly decaying solution $u_j=1/(j-1)!$. Yet, the
slowly decaying solution is unphysical because the sum $\sum_j u_j$
diverges. Therefore, the only possible solution is the trivial one,
$u_j=0$ for all $j\geq 1$, and we conclude that there are no
macroscopic hubs in the first generation.

Similarly, there are no macroscopic hubs in higher generations.  If
there is a hub in the $n$th generation, the recurrence \eqref{uj-rec}
holds for $j>n$ with the boundary condition $u_n=0$, or equivalently
$u_{n+1}=u_{n+2}$. By repeating the steps above, we conclude that all
coefficients $u_j$ vanish.

\section{Highly Connected Nodes}
\label{high}

In the previous section we have shown that the degree of the most
connected node in the first generation cannot be macroscopic. It is
then natural to assume that the degree scales as $N^{1/\beta}$ with
$\beta>1$, or equivalently \cite{OK}
\begin{equation}
\label{Hj-seek}
H_j=H_2\,h_j,\quad {\rm with}\quad H_2\simeq C\,N^{1/\beta},
\end{equation}
for all $j\geq 2$. The fractions $h_j$ satisfy
\begin{equation}
\label{hj-eq}
h_j=\beta\,\sum_{i\geq j} i^{-1}  h_{i-1}\,,
\end{equation}
obtained by substituting \eqref{Hj-seek} into \eqref{Hj}.  Therefore,
the recursion relation \eqref{gn-rec} becomes 
\begin{equation}
\label{hj-rec}
h_j=\beta\, j^{-1}h_{j-1}+h_{j+1}.
\end{equation}
The boundary conditions is
\begin{equation}
\label{hj-in}
h_1=0, \quad h_2=1.
\end{equation}

Certain features can be determined without solving
Eqs.~\eqref{hj-rec}--\eqref{hj-in}.  For instance, to obtain the total
number of descendants $h_{\rm tot}=\sum_{i\geq 2}h_i$ we multiply
\eqref{hj-rec} by $j$ and sum over all $j$. This yields
\begin{equation}
\label{htot}
h_{\rm tot}=\frac{1}{\beta-1}\,.
\end{equation}
Since $h_{\rm tot}>1$, we have the bound $\beta<2$.

Similarly, by multiplying \eqref{hj-rec} by $j^2$ and summing over all
$j$ we determine the average generation number
\begin{equation*}
\langle \,j\,\rangle=\frac{\sum_{j\geq 2} jh_j}{\sum_{j\geq 2} h_j} =
\frac{2}{2-\beta}\,.
\end{equation*}

The second order linear difference equation \eqref{hj-rec} has two
independent solutions. We obtain the asymptotic behavior of these
solutions by converting the difference equation \eqref{hj-rec} into a
differential equation. If $h_j$ decays sufficiently slow, we may
approximate Eq.~\eqref{hj-rec} by the differential equation
$dh/dj\simeq -\beta\,h/j$ with solution \hbox{$h_j\sim j^{-\beta}$} at
large $j$. Otherwise, if $h_j$ decays very quickly, we have $h_j\simeq
\beta\,h_{j-1}/j$ in the leading order and therefore $h_j\sim
\beta^{j}/j!$. Using this solution as an integrating factor, that is,
substituting $h_j=q_j\,\beta^{j}/j!$ into Eq.~\eqref{hj-rec}, we have
the recurrence $q_j-q_{j-1}=\beta q_{j+1}/(j+1)$.  This recurrence
reduces to the differential equation $dq/dj \simeq \beta\,q/j$, with
power-law solution $q_j\sim j^{\beta}$.  Therefore the asymptotic
behavior of the second solution to Eq.~\eqref{hj-rec} is
\begin{equation}
\label{hj-asym}
h_j\sim \frac{j^\beta\,\beta^j}{j!}.
\end{equation}

Equations \eqref{hj-rec}--\eqref{hj-in} admit a solution for {\em
arbitrary} value of $\beta$. In general, such a solution is a linear
combination of the two solutions above.  However, the slowly decaying
solution is unphysical. If $H_j\sim j^{-\beta} N^{1/\beta}$ for $j\gg
1$, then the depth of the subtree emanating from the most connected
node in the first generation, \hbox{$j_{\rm max}\sim N^{1/\beta^2}$},
as follows from the criterion $H_{j_{\rm max}}\sim 1$, would exceed
the maximal depth \eqref{n-max}.  Therefore, we must find the special
value of $\beta$ for which solution to \eqref{hj-rec}--\eqref{hj-in}
is the rapidly decaying solution with the asymptotic behavior
\eqref{hj-asym}. Therefore, $\beta$ is analogous to an eigenvalue.

To determine the proper eigenvalue $\beta$ we use the generating
function technique. We write
\begin{equation}
\label{hz-ref}
\mathcal{H}(z) = \sum_{j\geq 2} h_j z^j
\end{equation}
and convert the difference equation \eqref{hj-rec} into an integral
equation
\begin{equation}
\label{H-eq}
\mathcal{H}(z) = \beta \int_0^z dx\,\mathcal{H}(x) +
\frac{\mathcal{H}(z) -z^2}{z}.
\end{equation}
Note that $\mathcal{H}(z) = z^2 + z^3 +\cdots$ for small $z$ and in
addition, Eq.~\eqref{htot} sets $\mathcal{H}(1) = h_{\rm tot} =
(\beta-1)^{-1}$.

Given the form of Eq.~\eqref{H-eq}, we focus on the integral of the
generating function, $\mathcal{F}(z) = \int_0^z dx\,\mathcal{H}(x)$,
and convert the integral equation \eqref{H-eq} into the differential
equation
\begin{equation}
\label{F-eq}
\frac{d\mathcal{F}}{dz} +\beta\,z\,(1-z)^{-1}\mathcal{F} = z^2(1-z)^{-1}.
\end{equation}
The boundary condition is $\mathcal{F}(0)=0$.  We use the
integrating factor $I(z)=e^{-\beta\,z}(1-z)^{-\beta}$ to simplify
\eqref{F-eq} to $d(\mathcal{F}I)/dz=z^2(1-z)^{-1}I$.
Hence, the solution of \eqref{F-eq} is
\begin{equation}
\label{F-sol}
\mathcal{F}(z) = e^{\beta(z-1)}(1-z)^{\beta}
\int_0^z dx\, e^{\beta(1- x)} \, \frac{x^2}{(1-x)^{1+\beta}}\,.
\end{equation}

The behavior of the generating function in the limit $z\to 1$ reflects
the behavior of the coefficients $h_j$ in the limit
$j\to\infty$. Therefore, we evaluate the integral in \eqref{F-sol} in
the limit $z\to 1$,
\begin{eqnarray*}
&&\int_0^z dx\, e^{\beta(1-x)}\,x^2\,(1-x)^{-1-\beta}\\
&=&\int_0^z dx\, e^{\beta(1-x)}\left[1-2(1-x)+(1-x)^2\right](1-x)^{-1-\beta}\\
&=&\int_{1-z}^1 dy\,e^{\beta y}[ y^{-1-\beta}-2y^{-\beta}+y^{1-\beta}]\\
&=&\sum_{k=0}^\infty\frac{\beta^k}{k!}\int_{1-z}^1 dy\, \left[y^{k-1-\beta}-2y^{k-\beta}+y^{k+1-\beta}\right]\\
&\simeq&f(\beta)+(1-z)^{-\beta}\left[\frac{1}{\beta}-\frac{2-\beta}{\beta-1}(1-z)\right].
\end{eqnarray*}
The function $f(\beta)$ is given by the infinite sum
\begin{equation*}
f(\beta)=\sum_{k=0}^\infty\frac{\beta^k}{k!}
\left(\frac{1}{k-\beta}-\frac{2}{k+1-\beta}+\frac{1}{k+2-\beta}\right).
\end{equation*}
We can also express this function in terms of the incomplete Gamma
function $\gamma(a,x)=\int_0^x dt\, t^{a-1}\,e^{-t}$,
\begin{equation}
\label{fb-exp}
f(\beta)=\frac{(-\beta)^{\beta-1}\gamma(2-\beta,-\beta)
+e^{\beta}}{\beta(\beta-1)}.
\end{equation}

Using $\mathcal{H}=d\mathcal{F}/dz$ and the leading $z\to 1$
asymptotics of the integral in \eqref{F-sol} we find
\begin{equation}
\mathcal{H}(z)\simeq \frac{1}{\beta-1}-\beta
f(\beta)(1-z)^{\beta-1} \quad\text{as}\quad z\to 1.
\end{equation}
Generally, the generating function has a regular component and a
singular one, $\mathcal{H}(z)=\mathcal{H}_{\rm
  reg}(z)+\mathcal{H}_{\rm sing}(z)$, with
\begin{equation*}
\begin{split}
\mathcal{H}_{\rm reg}(z)  & \simeq \frac{1}{\beta-1}+{\rm const}\times (1-z),\\
\mathcal{H}_{\rm sing}(z) &\simeq -\beta\,f(\beta)(1-z)^{\beta-1}.
\end{split}
\end{equation*}
Since $\sum_j z^j j^{-\beta}\sim (1-z)^{\beta-1}$ when $z\to 1$, we
identify the singular component with the unphysical solution.  Hence,
the singular term must vanish, and the exponent $\beta$ is root of the
transcendental equation
\begin{equation}
\label{root}
f(\beta)=0.
\end{equation}
By solving this equation numerically, using the bisection method
\cite{nr} for example, we find the eigenvalue to essentially arbitrary
precision 
\begin{eqnarray*}
\beta=1.351746470331\ldots\,.
\end{eqnarray*}

The above behavior extends to higher generations. In general, the
degree of the most connected node at the $n$th generation grows
algebraically with the total number of nodes, $H_{n+1}\sim
N^{1/\beta_n}$, and the exponent $\beta_n$ is function of the
generation index $n$.  Let $H_j$ be the number of descendants of the
most connected node at generation $j>n$. The coefficients $h_j$, 
defined by $H_j=H_{n+1}h_j$, again satisfy the recursion \eqref{hj-eq}
with the boundary condition $h_n=0$ and $h_{n+1}=1$.  Thus, the
coefficients $h_j$ decay super-exponentially as in \eqref{hj-asym}.

The exponents $\beta_n$ are obtained by repeating the steps above, and
we merely quote the final results. First, the integral of the
generating function is
\begin{equation}
\label{F-sol-n}
\mathcal{F}(z) = e^{\beta_n z}(1-z)^{\beta_n}
\int_0^z dx\, e^{-\beta_n x} \, \frac{x^{n+1}}{(1-x)^{1+\beta_n}}\,.
\end{equation}
Second, the function $f(\beta)$ is now
\begin{equation}
\label{fbn}
f(\beta_n)=\sum_{k=0}^\infty\frac{\beta_n^k}{k!}
\sum_{m=0}^n {n\choose m}\frac{(-1)^m}{k+m-\beta_n}.
\end{equation}
The values of the exponents $\beta_n$ for $n\leq 8$ are listed in
table I.  The exponents increase with $n$ and therefore, the degree
of the most connected nodes declines sharply as distance from the root
increases.

\begin{table}[ht]
\begin{tabular}{|c|l|l|l|}
\hline
$n$&$\beta_n$&$h_{\rm tot}$&$\langle \Delta j\rangle$\\
\hline
$0$&$1$&$1$&2\\
$1$&$1.351746$&$2.8429$&$2.0852$\\
$2$&$1.682201$&$2.9316$&$2.1466$\\
$3$&$2$&$3$&$2.1945$\\
$4$&$2.309250$&$3.0551$&$2.2336$\\
$5$&$2.612266$&$3.1012$&$2.2665$\\
$6$&$2.910493$&$3.1405$&$2.2949$\\
$7$&$3.204901$&$3.1747$&$2.3197$\\
$8$&$3.496180$&$3.2048$&$2.3418$\\
\hline
\end{tabular}
\caption{The exponent $\beta_n$, the ratio $h_{\rm tot}$ between the
  total number of descendants and the degree, and the average distance
  of a descendant $\langle \Delta j\rangle$ versus generation $n$. The
  exponents $\beta_n$ are roots of the transcendental equation
  $f(\beta_n)=0$ with $f(\beta_n)$ given by \eqref{fbn}. The quantities
  $h_{\rm tot}$ and $\langle \Delta j\rangle$ are given in equations
  \eqref{htot-n} and \eqref{delta-n}, respectively.}
\end{table}

The expression \eqref{fbn} holds for all $n$ as long as $\beta_n$ is
noninteger.  There are, however, two exceptions.  We have already seen
that for the zeroth generation $\beta_0=1$ and the coefficients $h_j$
can be determined analytically, $h_j=1/(j-1)!$. For the third
generation, $\beta_3=2$ and again, the coefficients $h_j$ have simple
form
\begin{equation}
\label{hj-sol3}
h_j = 3\cdot 2^{j-3}\,\frac{j-3}{(j-1)!}.
\end{equation}
This can be verified by direct substitution into \eqref{hj-rec}. Of
course, the solution \eqref{hj-sol3} agrees with the general
asymptotic behavior \eqref{hj-asym}.

Using $\beta_n$, we calculate several other useful properties.
First, the ratio $h_{\rm tot}$ between the total number of descendants
and the degree equals
\begin{equation}
\label{htot-n}
h_{\rm tot}=\frac{n}{\beta_n-1}.
\end{equation}
This quantity mildly increases with generation and eventually
saturates (table 1).  Also, we mention the average distance of a
descendant $\langle \Delta j\rangle = \langle j\rangle -n$ with
$\langle j\rangle =\sum_{j\geq n} jh_j/\sum_{j\geq n} h_j$.  This
quantity is given by
\begin{equation}
\label{delta-n}
\langle \Delta j\rangle=\frac{n-\beta_n-1}{\beta_n-2}.
\end{equation}
When $n=3$, $\langle \Delta j\rangle=(e^2-3)/2$. Like $h_{\rm tot}$,
the average distance slowly grows with generation but ultimately
approaches a constant. Hence, statistical properties of average
quantities such as the total number of descendants and the distance
to a descendant weakly depend on depth, but statistical properties of
extreme quantities such as the degree of the most connected node
strongly depend on depth.

\section{The Degree Distribution}
\label{degree}

\begin{figure}[t]
\includegraphics[width=0.45\textwidth]{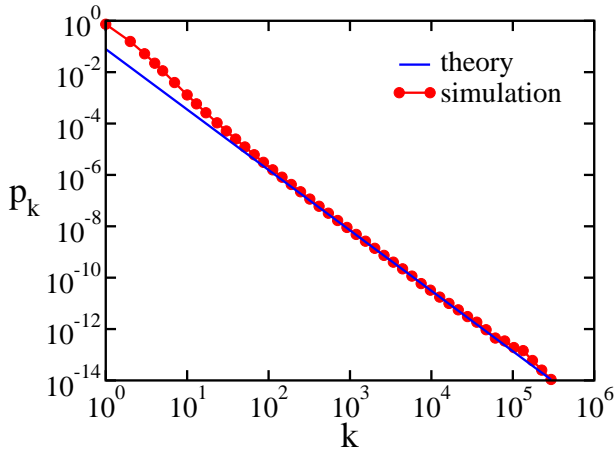}
\caption{The degree distribution $p_k$ versus degree $k$. The
simulation results (bullets) are from $10^2$ independent realizations
with $N=10^8$. Also shown is a straight line that corresponds to the
theoretical result \eqref{announce}.}
\label{fig-pk}
\end{figure}

The degree distribution is an important quantity that characterizes
local properties of networks. For many growing networks, the degree
distribution satisfies closed recurrence equations and consequently,
this quantity can be determined analytically.  It appears impossible
to find a closed set of equations that govern the degree distribution
for the random ancestor tree. Remarkably, the closed set of equations
\eqref{Hj} allows us to obtain the tail of the degree distribution
analytically.

The power-law degree of the most connected node \eqref{Hj-seek} already
shows that the degree distribution has power-law tail. Let $p_k$ be the
fraction of all nodes with degree $k$. As announced in
\eqref{announce}, this distribution decays algebraically,
\begin{equation}
\label{pk}
p_k\sim k^{-\gamma}\quad {\rm with}\quad\gamma=\beta+1, 
\end{equation}
for $k \gg 1$.  Indeed, if we substitute $k_{\rm max}\sim N^{1/\beta}$
with $\beta\equiv \beta_1$ into the extreme statistics criterion,
\hbox{$N\sum_{k>k_{\rm max}} p_k\sim 1$}, we obtain \eqref{pk}. Here,
we take $\beta= \beta_1$ because the exponents $\beta_n$ increase
monotonically.

Our numerical simulations results are in excellent agreement with the
theoretical prediction \eqref{announce}.  The simulations are
straightforward to implement: with each new node, a target node is
selected at random, and then, the attachment node is selected at
random from the target node and all of its ancestors.  Since the
average depth remains finite, the computational cost is proportional
to the network size and we can simulate networks as large as $N=10^8$.
We comment that many real-world networks have power-law
degree distributions \cite{ab} as in \eqref{pk} with exponent $\gamma$
in the range $2<\gamma<3$. For example, for the Internet at the
Autonomous Systems level, as of 2007, the exponent is approximately
$\gamma=2.5$, see \cite{cskss}.

\begin{figure}[t]
\includegraphics[width=0.45\textwidth]{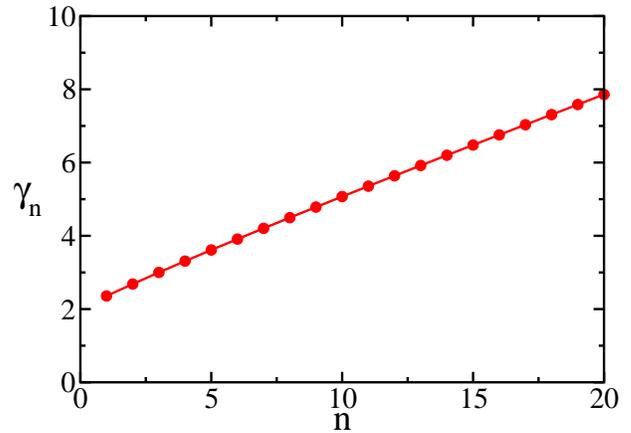}
\caption{The exponent $\gamma_n=1+\beta_n$ versus $n$. }
\label{fig-gn}
\end{figure}

Remarkably, the random ancestor tree has a family of degree
distributions, each characterized by a distinct nontrivial
exponent.  The fraction $p_{n,k}$ of $n$th generation nodes with
degree $k$ decays algebraically
\begin{equation}
p_{n,k}\sim k^{-\gamma_n}\quad {\rm with}\quad\gamma_n=\beta_n+1.
\end{equation}
As in the preferential attachment network, the exponent that
characterizes the degree distribution increases with generation
\cite{bk1}.

From the roots of the transcendental equation \hbox{$f(\beta_n)=0$}
with $f(\beta_n)$ given by \eqref{fbn}, we find that the exponent
$\gamma_n$ grows linearly with $n$, that is (figure \ref{fig-gn})
\begin{equation}
\gamma_n\sim n.
\end{equation}
Since $p_k=\sum_{n\geq 1} g_n p_{n,k}$ and since $g_n$ increases with
$n$, the dominant contribution to the distribution $p_k$ comes from
first generation nodes. Hence, $\gamma\equiv\gamma_1$. This behavior
is in contrast with the preferential attachment network where the
behavior at the average generation dominates the degree distribution
\cite{bk1}.

For completeness, we mention that the in-component distribution mimics
the degree distribution. The degree of the in-component of a node is
defined as the total number of its descendants. The quantity $h_{\rm
  tot}$ above is always finite and hence, the total number of
descendants of the most connected nodes is proportional to the
degree. As consequence, the fraction $I_s$ of nodes with in-component
degree $s$ decays algebraically, $I_s\sim s^{-\gamma}$.

\section{Leaves, Stars, and Chains}
\label{leaves}

It is also possible to calculate a few other structural properties of
the network including the average number of leaves as well as the
probabilities that the network has an extreme structure such as a star
or a chain.

Nodes with no incoming links, or equivalently, nodes with degree one
are termed leaves. To calculate the total number of leaves, we must
first find a more detailed quantity: $L_n(N)$, the total number of
leaves at the $j$th generation. This quantity evolves as follows
\begin{equation}
\label{Ln-eq}
\frac{dL_n}{dN} = \sum_{i\geq n} \frac{G_{i-1}}{i\,N} -
\frac{L_n}{(n+1)N}.
\end{equation}
Each new node is necessarily a leaf and hence, the gain term is as in
\eqref{Gn}.  The loss term reflects the fact that any new link into a
leaf decreases the number of leaves by one.

We assume that the number of leaves in each generation is macroscopic,
\begin{equation}
\label{Ln-seek}
L_n(N) = \ell_n N.
\end{equation}
By substituting this form into the rate equation \eqref{Ln-eq}, we
find that leaves constitute a finite fraction of all nodes in each
generation, $\ell_n = \frac{n+1}{n+2}\,g_n $, or equivalently,
\begin{equation}
\label{Ln-sol}
\ell_n = e^{-1}\,\frac{n(n+1)^2}{(n+2)!}.
\end{equation}
The fraction of leaves grows with depth: 2/3 of all nodes in the first
generation are leaves; 3/4 of all nodes in the second generation are
leaves; etc.  The total number of leaves, $L_{\rm tot}$, is the sum
$L_{\rm tot}=\sum_{n\geq 1} L_n$ and from \eqref {Ln-sol}, we obtain
\begin{equation}
\label{ltot}
L_{\rm tot}=2e^{-1}N.
\end{equation}
As is often the case in complex networks, a large fraction of all
nodes have no incoming links.

It is also possible to calculate the likelihood that the tree has one
of two extreme topologies: (i) a star where all nodes link to the
root, and (ii) a linear chain of length $N$. Let $S_N$ be the
probability that the tree is a star graph.  This probability satisfies
the recursion
\begin{equation}
S_{N+1}= S_N\left(\frac{1}{N}+ \frac{N-1}{2N}\right).
\end{equation}
The network is a star graph only if it was previously a star.
Furthermore, the factor $1/N$ is the probability that the root is the
target node and the factor $(N-1)/(2N)$ is the probability that the
target node is any one of the $N-1$ leaves, but the actual link is
made to the root. With the boundary condition $S_1=S_2=1$, the
probability $S_N$ decays exponentially with the network size,
\begin{equation}
\label{SN}
S_N= N\,2^{-(N-1)}.
\end{equation}

The likelihood that the tree has a chain topology, $C_N$, obeys the
recursion equation $C_{N+1}=C_N/N^2$. This recursion reflects that the
newest node must always be selected both as the target node and as the
attachment node. The probability for this event is
$1/N^2$. Using the boundary condition $C_2=C_3=1$, the likelihood of
growing a chain decays extremely rapidly,
\begin{equation}
\label{CN}
C_N= \frac{1}{[(N-1)!]^2}.
\end{equation}
Therefore, it is far more likely that the random ancestor tree is a
star than it is a chain. This is another consequence of the shallow
nature of the tree.

\section{Conclusions}
\label{conclude}

We introduced a random structure where the genealogy governs the
evolution. The random ancestor tree has a remarkably rich structure.
The network is very tight with a sharply-decaying distribution of
depth. There is a single macroscopic hub that is connected to a finite
fraction of all nodes along with multiple highly connected nodes.

The network is strongly stratified because the genealogical structure
controls the growth.  The most connected node at distance $n$ from the
hub has a sub-macroscopic degree that scales as $N^{1/\beta_n}$ with
system size $N$. Interestingly, the exponents $\beta_n$ are generally
transcendental numbers.  Moreover, the exponents $\beta_n$ grow
monotonically with $n$, and thus, the connectivity sharply declines
with increasing depth.  As a consequence, the degree distribution has
power-law tail, $p_k \sim k^{-\gamma}$ with $\gamma=\beta_1+1$ and
furthermore, the degree distribution varies strongly with depth.

We obtained all of these features analytically using the total number
of descendants of a highly connected node as a function of
generation. This quantity obeys a closed set of equations and it
allows us to determine many scaling properties including in
particular, the exponent that governs the tail of the degree
distribution. This theoretical technique has promise in other growing
trees problems when the degree distribution does not satisfy closed
equations \cite{bk,mgn}, especially in situations where the
distribution of depth becomes independent of system size
asymptotically.

The random ancestor tree includes no control parameters but can be
easily generalized by various modifications of the attachment process.
More generally, the target node can be selected at a rate that is
proportional to the degree and similarly, the attachment node can be
determined according to either the generation number or the degree. We
envision that such generalizations can be useful for controlling the
degree distribution or the number of macroscopic hubs and hence,
relevant for modeling complex networks. The most challenging
generalization is the theoretical understanding of ancestor networks
with cycles.

\acknowledgments We thank Hasan Guclu, Renaud Lambiotte, and Sidney
Redner for useful discussions.  We are grateful for financial support
from DOE grant DE-AC52-06NA25396 and NSF grant CCF-0829541. PLK thanks
the Theoretical Division and the Center for Nonlinear Studies at Los
Alamos National Laboratory for hospitality.


\begin{thebibliography}{99}
\bibitem{OK} The ansatz \eqref{Hj-seek} is consistent with the
  analysis of Sec.~\ref{hubs} because $u_j\sim N^{1/\beta-1}$
  vanishes in the thermodynamic limit.  The amplitude $C$ that appears
  in \eqref{Hj-seek} varies from realization to realization, but
  nevertheless, the fractions $h_j$ become deterministic in the
  thermodynamic limit.
\end{thebibliography}
\end{document}